# Epsilon near zero metal oxide based spectrally selective reflectors

*Sraboni Dey\*, Kirandas P S, Deepshikha Jaiswal Nagar and Joy Mitra\**

School of Physics, Indian Institute of Science Education and Research Thiruvananthapuram, Kerala - 695551, India

\*Email: srabonidey20@iisertvm.ac.in,  j.mitra@iisertvm.ac.in

ABSTRACT

Epsilon near zero (ENZ) materials can contribute significantly to the advancement of spectrally selective coatings aimed at enhancing efficient use of solar radiation and thermal energy management. Here, we demonstrate a subwavelength thick, multilayer optical coating that imparts a spectrally "step function" like reflectivity onto diverse surfaces, from stainless steel to glass, employing indium tin oxide as the key ENZ material. The coating, harnessing the ENZ and plasmonic properties of nominally nanostructured ITO along with ultrathin layers of Cr and $Cr_2O_3$ show 15% reflectivity over the visible to near-infrared and 80% reflectivity (and low emissivity) beyond a cut-in wavelength around 1500 nm, which is tunable in the infrared. A combination of simulations and experimental results are used to optimize the coating architecture and gain insights into the relevance of the components. The straightforward design with high thermal stability will find applications requiring passive cooling.

## 1. INTRODUCTION

Solar radiation management, a key component of sustainable energy generation, is an area of wide topical interest wherein spectrally selective coatings form an important domain of research [1, 2]. Engineering spectral selectivity of surfaces through coatings offers an opportune way to maximize effective utilization of incident and radiated energies. This discussion focuses on realizing a coating exhibiting idealized step-function-like reflectivity (R), characterized by high absorption in the visible and high reflectance and low emissivity in the infrared, separated by a cut-in wavelength $\lambda_o$. Design and fabrication of such coatings that are durable, cost-effective and scalable, ensuring compatibility with diverse surfaces remain open challenges. Strategies for inducing such selectivity have broadly centered on assembling the right materials (semiconductors, metals and dielectrics) in conducive form factors like multilayers or composites with two and three-dimensional texturing [3, 4]. The increasing ability to engineer hybrid materials with sub-wavelength scale layering [5] and structuring [6] has broadened the scope of tailoring spectral selectivity. Fundamentally, the above schemes rely on the physics of light-matter interactions and engineering radiative vs. non-radiative modes, as realized in photonic crystals [7, 8] and optical metamaterials [9, 10]. Interestingly, non-radiative regimes are also realized in continuous media such as the epsilon near zero (ENZ) materials, near their ENZ wavelength ($\lambda_{ENZ}$).

Conducting oxides like Al-doped ZnO, indium tin oxide, doped CdO etc. [11, 12] harbor ENZ regimes that originate from their free electron response. As $\lambda \to \lambda_{ENZ}$, the real part of their relative permittivity ($\epsilon(\omega) = \epsilon' + i\epsilon''$), $\epsilon' \to 0$ and the $\lambda_{ENZ}$ demarcates a dielectric to metallic transition in terms of their optical properties. Consequently, ENZ systems sustain non-radiative modes around $\lambda_{ENZ}$ and ultra-thin layers (t ~ 10 nm) show perfect absorption (PA) with strong directional preference [13]. Spectrally, PA is realized at wavelengths $\lambda_{PA} \lesssim \lambda_{ENZ}$ and is coupled with excitation of the ENZ modes accompanied by extreme field amplification and confinement [14]. Further, the metallic response of the ENZ thin films for $\lambda > \lambda_{ENZ}$, ensures high reflectivity ($R \to 1$) at longer wavelengths, which together with $R \sim 0$ for PA enables a niche application of ENZ thin films in developing spectrally selective coatings of sub-wavelength dimensions.

Coatings that impart a step-function-like reflectivity (SFR) to flat surfaces are of vital interest in solar energy conversion[15,16], thermal management[17,18], radiative cooling[19], smart windows [20] and related areas. The optical properties of surfaces like stainless steel (SS), Si, glass, Cu etc. that are used across diverse platforms have been variously modified to impart SFR, as summarized in table 1. Typically, the samples (coated substrate) are dominantly absorbing (or transmissive) in the visible to near-infrared ($\lambda < \lambda_o$) and become highly reflecting above $\lambda_o$. The latter minimizes radiative energy loss by minimizing thermal emissivity, for $\lambda > \lambda_o$, and thus a sharper transition from low to high $R$ across the $\lambda_o$ increases the overall efficiency of the coating, in terms of thermal emissivity.

Table 1: Details of coatings that impart spectrally selective and step function like variation in Reflectivity on various surfaces.

| Material on Substrate | Absorptivity (VIS/NIR) | Emissivity (E) (IR) | Sharpness of transition ($\Delta\lambda$) (nm) | Refs. |
|---|---|---|---|---|
| Cr/SiO₂/Cr on Si | 0.99 | - | - | [21] |
| W/WAlN/WAlON/Al₂O₃ on SS | 0.90 | 0.15 | ~2000 | [22] [23] |
| TiAlC/TiAlCN/TiAlSiCN/TiAlSiCO/TiAlSiO on SS | 0.961 | 0.07 | ~1000 | [24] |
| 3 -SiC-W nanocomposite layers on Si | 0.95 | 0.05 | ~3000 | [25] |
| Structured graphene metamaterial on Cu * | ~0.90 | ~0.04 | ~1800 | [26] |
| ITO NS /ITO / Cr/ Cr₂O₃ on SS and Glass * | 0.87 (SS) | 0.18(SS) | ~400 | This work |
| Cu/Al₂O₃/Cr/SiO₂/Cr/SiO₂ on Glass and Si | 0.954 | 0.196 | ~2000 | [27] |
| Au/NiCr–MgF₂ (HMVF)/ NiCr–MgF₂ (LMVF)/MgF₂ on SS | 0.976 | 0.045 | ~3500 | [16] |
| Multilayer graphene + TiN based metamaterial * | 0.88 | 0.03 | ~2000 | [28] |
| $SiO_2$ / $Al_2O_3$/ $ZrB_2$/ $Al_2O_3$/ $ZrB_2$/ $ZrB_2$/ $Al_2O_3$ / $ZrB_2$ on SS and Si | 0.96 | 0.16 | ~2000 | [29] |
| W/WAlSiN/SiON/$SiO_2$ on SS and Si | 0.955 | 0.10 | ~2000 | [30] |
| Multilayer reduced graphene oxide on Al * | 0.92 | 0.04 | ~1,750 | [31] |
| CrN(H)/CrN(L)/CrON/$Al_2O_3$ on SS | 0.93 | 0.14 | ~3000 | [32] |
| Pt/ TiO2/ Al on SS * | 0.90 | - | - | [33] |
| Black chrome/ITO/$SiO_2$ on SS | 0.90 | 0.4 | ~1200 | [34] |

*Tunability demonstrated

However, these coatings involve significant complexity in growth, optimization and incorporation into a coating format, and a majority of their spectral response exhibits a wide low to high $R$ crossover regime, $\Delta\lambda \gtrsim 2\ \mu m$. Thus narrowing $\Delta\lambda$ to lower emissivity, improving stability and spectral tunability are challenges that form the basis of ongoing research in the area.

Leveraging the novel optical properties of ENZ materials like ITO, the primary aim here has been to demonstrate (i) SFR on opaque surfaces like SS, (ii) wide-angle SFR response and be of (iii) sub-wavelength thickness with nominal nano-structuring and for performance comparison, the same coating has been tested on a transparent material like glass. The design of meta-surfaces, broadband absorbers, and spectrally selective absorbers and reflectors using ITO have been studied in recent times [34, 35, 36, 37, 38]. Importantly, ITO has shown significant advantages compared to other transparent conducting oxides (TCOs) owing to its robustness in moist air, thermal stability, surface smoothness, mechanical stability and simple fabrication process [39] thereby encouraging its usage in spectral selective optical systems. Experimentally measured optical and thermal properties of coated surfaces, in conjunction with simulated responses, are used to comprehend the physics of surface engineering, where an ENZ material (ITO) imparts spectrally selective, wide-angle optical properties to commercially relevant substrates like opaque SS and glass from the visible to IR. The developed coating is a multi-layer of thin films of ITO/Cr/$Cr_2O_3$/substrate. Further, an elementary nanostructured grating of ITO (Fig 1) is shown to further tune $\lambda_o$ beyond the $\lambda_{ENZ}$ of the ENZ material. The ( wide angle 0° - 60°), "step function" like spectral response is demonstrated for both flat and nanostructured coatings for $\lambda_o$ around 1500 nm along with a sharp $\Delta\lambda \sim 400$ nm. Finally, control of $\lambda_o$ via changing the free electron density of ITO is demonstrated. The coating structure is readily adapted as a traditional solar absorber for $\lambda_o \sim 2500$ nm and the concept is indeed functional in other spectral ranges with relevant ENZ materials.

## 2. RESULTS AND DISCUSSION

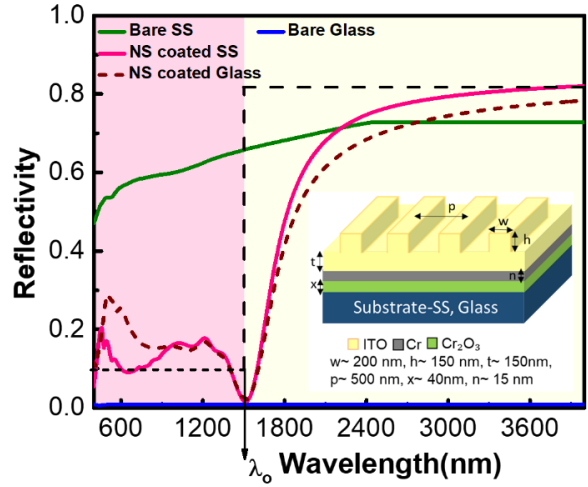

Figure 1. Spectral reflectivity showing the ideal "step function" like response (dashed line) along with the simulated response for bare SS and glass and with nanostructured coating. Inset shows schematic of the nanostructured coating.

Fig. 1 shows the schematic of the multilayer coating with NS, and the ideal SFR (dashed line) along with $R$ for bare SS and glass and those simulated using $p$-polarized light with NS coating on both substrates, averaged over incident angles from $0° - 60°$. Bare SS shows a monotonic variation in $R$ from $0.5 - 0.7$, while glass displays complete transparency. The NS coating radically changes the reflectivity of both the opaque and transparent substrate with their simulated $R$ vs. $\lambda$ displaying SFR response.

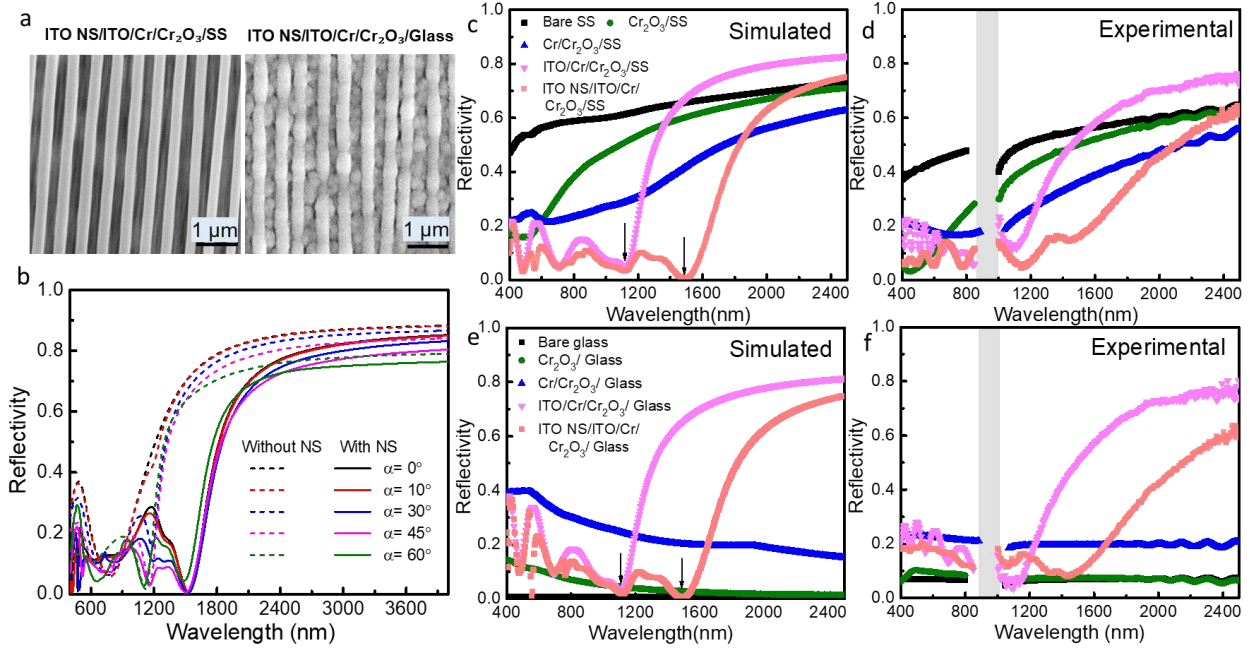

Figure 2(a) SEM images of ITO NS/ITO/Cr/Cr$_2$O$_3$ on SS and glass; (b) Simulated reflectivity spectra for SS coated with and without NS coating at selected angles for $\lambda_{ENZ} = 1220\ nm$; Simulated and Experimental reflectivity spectra of- (c)(d) Bare SS, Cr$_2$O$_3$ on SS, Cr/ Cr$_2$O$_3$/SS, ITO/Cr/ Cr$_2$O$_3$/SS, ITO NS/ITO/ Cr/ Cr$_2$O$_3$/SS; (d)(e) Bare BK7 glass, Cr$_2$O$_3$ on Glass, Cr/ Cr$_2$O$_3$/Glass, ITO/Cr/Cr$_2$O$_3$/Glass, ITO NS/ ITO/ Cr/ Cr$_2$O$_3$/Glass. Down arrows denote the $\lambda_o$ with and without NS. The shaded region in experimental plots-noise from the spectrometer.

Fig. 2a shows the secondary electron images of the coated substrates, showing the grating NS on each. Morphological features of the bare substrates and after deposition of each layer are shown in SI Section 2 (figs. S3 and S4), with the latter demonstrating large area patterning of the substrates. Figs. 2(c-d) show the simulated and experimental reflectivity spectra for SS as it is coated by the various components of the multilayer, and figs. 2(e-f) show the same for glass – recording evolution of reflectivity. They demonstrate quantifiable agreement between the simulated and experimental results and elucidate the role of individual layers in realizing the final response. Note that coatings with flat ITO and NS ITO, both show the SFR response with different $\lambda_o$. In all cases the $\lambda_o$ is taken to be the wavelength with the smallest reflectivity before transition. ITO is a low-loss dielectric for $\lambda < \lambda_{ENZ}$, thus the low R at shorter $\lambda$ (fig 1) arise primarily from the optical properties of the underlying layers Cr/Cr$_2$O$_3$/substrate. On SS, this structure realizes a Fabry Perot cavity for *p*-polarized light at wavelengths below 500 nm, with the reflective Cr and SS increasing the effective interaction of light with these lossy materials, which contributes to overall lowering of $R$. The low loss, non-dispersive nature of Cr$_2$O$_3$ from 400 – 4000 nm [40] allows tuning the absorption by optimizing the thickness of the Cr layer. The Cr layer with optimized thickness of 15 $nm$ is critically responsible for balancing the trade-off between light penetration and absorption, while ensuring its integrity as a uniform optical layer. ITO undergoes a dielectric to metal transition at $\lambda_{ENZ}$ and progressively becomes more reflective for $\lambda > \lambda_{ENZ}$ resulting in the high $R$ at longer $\lambda$. For wavelengths just below $\lambda_{ENZ}$, in the ENZ regime (where $(\epsilon') > 0$ and $n < 1$) ITO thin films also exhibit perfect absorption (PA) thereby reducing its $R \sim 0$. Thus, reflectivity for the coating without NS (fig 2b) starts increasing for $\lambda > \lambda_{ENZ}$ with $\lambda_o \approx \lambda_{ENZ}$. The relevance of the NS ITO grating is evidenced in the simulated plots of $R$ shown in fig. 2b. Both with and without NS the $R$ saturates to ~ 80% above 2000 nm however for coatings with NS, $\lambda_o \sim 1500\ nm$ is red-shifted from $\lambda_{ENZ}$ and is determined by the grating dimensions. The final dimensions of each element are optimized via simulations to deliver the desired spectral response as discussed in Section 1 of SI.

In spite of the correlation between the experimentally measured reflectivity and the simulations some major differences are apparent. The measured $R$ is systematically lower than the simulated value, across each layer and the $\Delta\lambda$ observed experimentally is higher than that predicted in the simulations. These departures primarily originate from scattering due to roughness of each of the deposited films and finally structuring.  Importantly, though the predicted and observed values of $\lambda_o$ for coatings with and without NS are in agreement with each other.  It is also interesting that the coatings render comparable reflectivity to diverse substrates like glass and SS, which highlights the determining role played by the multilayer. The glass coated substrate shows low reflectivity and transmittivity (<20%, see SI Section 3,fig. S7) in the visible for $\theta_{in} = 0° - 90°$, which benefits from the refractive index contrast of 1.5 (glass) and 2.2 ($Cr_2O_3$). It also enables total internal reflection for $\theta_{in} > 42°$ thus, increases interaction with the lossy Cr layer leading to absorption and suppressing reflectivity in the visible regime over a wide angular regime. Inspite of having low reflectivity in the visible , the nanostructured portions show structural colours as evident in the optical images around 45º angles (SI section 4, fig S8).

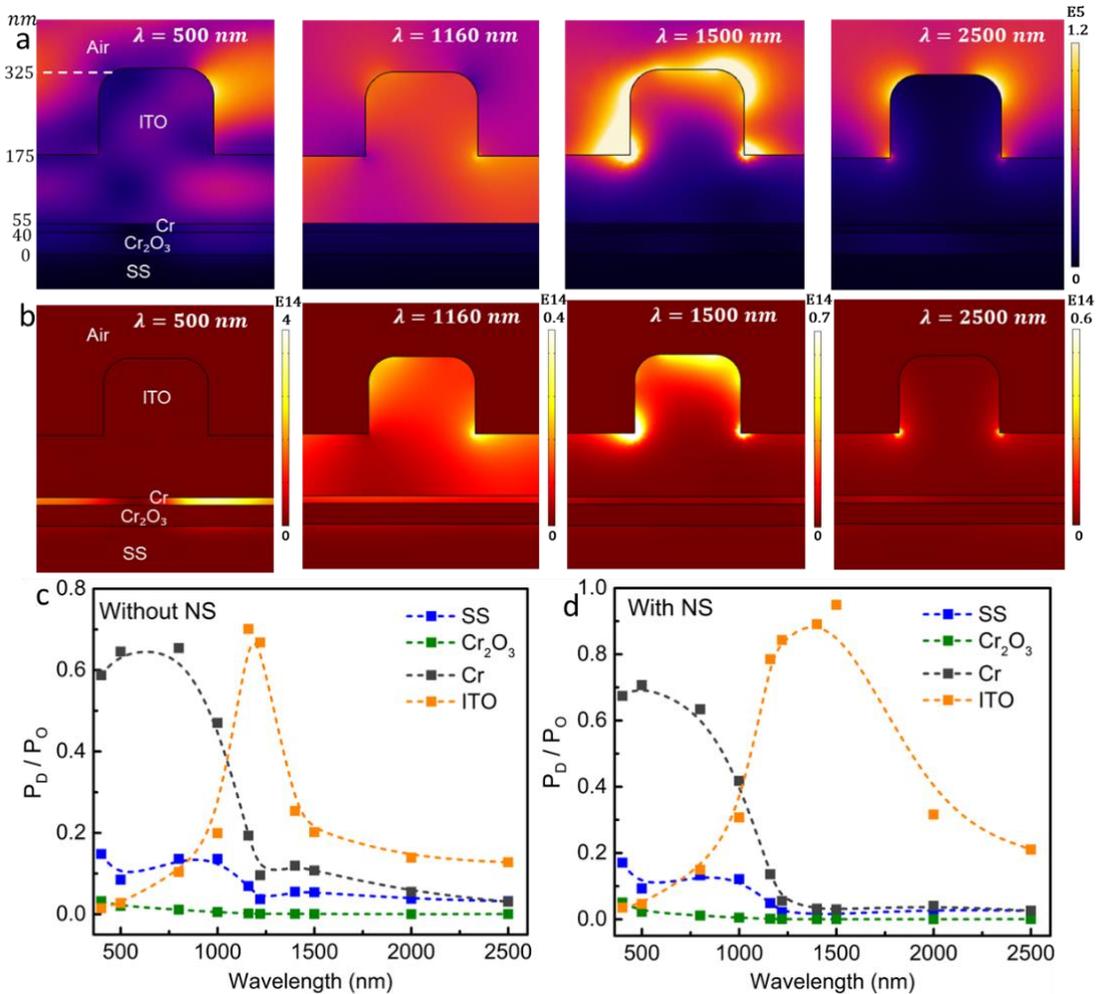

Fig 3. (a) Simulated electric field (V/m) and (b) dissipated power density (W/m³) across the system at selected wavelengths; Spectral variation of power dissipated as a fraction of incident power, across the various layers (c) without nanostructures and (d) with nanostructures.

The simulations offer further insights into the role of each component of the multilayer coating in eliciting the overall response. Figs. 3(a-b) plot the calculated electric field (|**E**|) and power dissipation density ($P_D = 1/2(\omega\varepsilon''|\mathbf{E}|^2)$) along the cross section of the NS coating on SS, at selected wavelengths. Figs. 3(c-d) show the spectral variation of net power dissipated ($\int P_D \, dv$), in the various layers for coatings with and without NS. For $\lambda < \lambda_{ENZ}$, $P_D$ is highest (~ 70%) in the Cr layer, the remaining radiation being absorbed in SS. Dissipation in ITO increases as $\lambda \rightarrow \lambda_{ENZ}$ and is highest around $\lambda_{ENZ}$ since $\epsilon' \rightarrow 0$, which maximizes the **E** field in ITO and thus $P_D$. Dissipation decreases at longer $\lambda$ due to the increasing reflectivity of ITO thus decreasing field penetration. For the coating with the NS $\lambda_o$ red-shifts to 1500 nm, around 300 nm beyond $\lambda_{ENZ}$ and dissipation remains high in ITO between $\lambda_{ENZ} - \lambda_o$. The **E** field plots 1500 nm show signatures of dipolar resonances originating from the plasmonic response of the ITO NS, dimensions of which are comparable to the $\lambda$. Also, the calculated **E** field is enhanced in the vicinity of NS compared to the flat coating as shown in SI section 5 Fig S9, indicative of the resonances in NS. They cumulatively impart high absorbance to the NS ITO coating between $\lambda_{ENZ} - \lambda_o$ and are responsible for red-shifting $\lambda_o$ showing that the narrow $\Delta\lambda \sim 400 \, nm$ transition width across the wide incidence angle is achievable only with the NS. Note that the reflectivity is also reduced in the visible with the introduction of the NS, as validated by the experimental results in fig. 2d and fig. 2f. The above features demonstrate the dominant role played by the properties of ITO in achieving the SFR response and contextualize the relevance of the NS.

The $\lambda_{ENZ}$ of ITO which crucially controls the $\lambda_o$ of the coatings is determined primarily by the electron density ($N_e$) of ITO, thus varying $N_e$ systematically changes $\lambda_{ENZ}$ [41,42,43,44] and $\lambda_o$. $N_e$ is readily tuned in ITO by altering the density of oxygen vacancies ($N_{VO}$) via annealing in various $O_2$ partial pressures [41]. Annealing in the ambient decreases $N_{VO}$ and consequently $N_e$ which red-shifts the $\lambda_{ENZ}$ and $\lambda_o$ as shown in fig. 4a and fig. 4c. Here, both for SS and glass substrates the reflectance spectra of the as deposited coating exhibits $\lambda_o \sim 1220\ nm$, which red-shifts from 1220 nm – 1580 nm upon annealing at $350°C$ for 10 mins as tabulated in SI section 6, Table S1 and S2. Fig. 4b and fig. 4d show the corresponding simulated spectra keeping all material parameters unchanged except decreasing $N_e$, which decreases from $1.05 \times 10^{21}$/cc to $6.62 \times 10^{20}$/cc. These results show that the coating is tunable yet thermally stable up to temperatures like 300 ºC. The calculated emissivity for the NS coated SS substrate, obtained by integrating the spectral response shows a low value of 0.2 in the IR with wide angular tolerance, along with high absorptivity ~ 0.8 over the visible to NIR, as discussed in SI section 7. The emissivity ~ 0.2, calculated over the IR ($\lambda > \lambda_o$) is indicative of the coating's efficiency in minimizing the radiative energy loss in the IR.

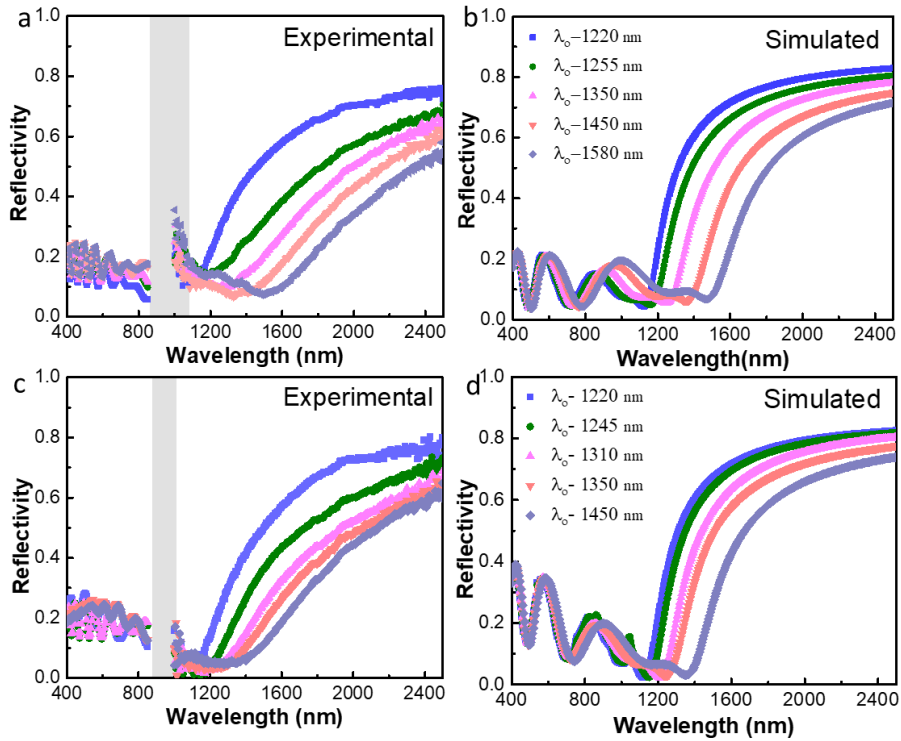

Figure 4. Experimental and Simulated reflectivity spectra of (a)(b) ITO/ Cr/ $Cr_2O_3$/SS; (c)(d) ITO/ Cr/ $Cr_2O_3$/Glass on annealing over certain time periods at 350 ºC. Shaded region in experimental plots-noise from the spectrometer.

The novelty of the coating's design and applicability in imparting the step function like reflectivity onto diverse surfaces is elucidated by the plot in SI fig. S11, which shows a SFR response (simulated $R\ vs.\ \lambda$) even for a free hanging coating i.e. with vacuum as the substrate. It demonstrates that the bulk of the response arises solely due to the coating and is minimally dependent on the optical properties of the substrate. However, the morphology of the substrate is still of relevance which needs to be optically flat to enable the grating nanostructure to perform as predicted in the IR.

3. **Conclusion**

In summary, we have developed a wide-angle, spectrally selective reflector coating that imparts step function like reflectivity on diverse substrates like SS and glass based on a straightforward design consisting of a tri-layer subwavelength thick $Cr_2O_3$, Cr and ENZ thin film of ITO and a nanostructured grating. The average reflectivity over the entire visible and NIR $\lambda < 1500$ nm is $\sim 15\%$ and thereafter becomes highly reflecting with reflectivity $\sim 80\%$. A combination of optical characterization and numerical simulations have been used to examine the role of individual layers in the integrated coating. Spectral variation in the calculated power dissipation across the layers elucidates the role of each layer, demonstrating the confinement of the electric field in the SS-$Cr_2O_3$-Cr layers, which uniformly suppresses reflectivity and maximizes absorption in the visible to NIR regime. The epsilon near zero property of ITO is exploited not only to impart strong reflectivity in the IR, but its nanostructured morphology is shown to improve the step function like reflectivity. Finally, the easy tunability of ITO's $\lambda_{ENZ}$ via thermal annealing is shown to systematically shift the overall spectral response of the coating in the IR and this is perhaps the first demonstration of spectral tunability of a coating utilizing an ENZ material. Such a mechanism of obtaining tunable spectral selectivity over optically diverse substrates (opaque SS and transparent glass) is promising in the field of energy management and demonstrates that the design ideas may be translated to other ENZ materials across other spectral regimes. Finally, this CMOS-compatible fabrication based on thin-film coatings and elemental nanostructuring is capable of achieving wide-angle, spectrally selective behavior offering a robust and cost-effective solution in optical engineering.

## 4. Materials and Methods

The multilayer structures were optimized in terms of the material properties and physical dimensions using finite element method modelling with 3D simulations conducted in the Wave Optics Module of COMSOL® Multiphysics 5.3a. The optimization scheme yielded the best optical response of the system considering feasibility of fabrication and experimental verification of the samples. It minimizes reflectance over the visible and maximizes reflectance above the cut-in wavelength ($\lambda_o$) with a straightforward fabrication protocol. The simulations were carried out in the range 400 – 4000 nm though the experimental verification is limited to 2500 nm. The optical properties of the materials were taken from the literature [40,45,46], with the Drude model used to determine that of ITO as a function of three parameters, carrier concentration $N_e$, scattering parameter γ and back ground permittivity $\epsilon_\infty$. The typical values of the parameters ($N_e$, γ and $\epsilon_\infty$) follow those from a previous publication [41] from the group. The details of the model geometry and its optimization are given in the supporting information (section 1).

**Sample fabrication**

Polished stainless steel (SS) substrates (~7 mm x 7 mm x 0.5 mm, MTI, USA) and NBK7 glass substrates (~7 mm x 7 mm x 1 mm) were cleaned thoroughly using acetone, isopropyl alcohol and finally with DI water. 40 $nm$ Cr was thermally evaporated on to the substrates at ~ 0.5 Å/s. The films were annealed at 600°C for 2 hours in the ambient to form $Cr_2O_3$, verified by XRD, EDX spectroscopy (EDS) and Raman spectroscopy, as shown in SI section 2. Subsequently, the samples were coated with 15 nm Cr (thermal evaporation) followed by coating ITO (RF sputtering) of variable thickness from 150 – 350 nm. Chamber pressure for all depositions were maintained at lower than 2×10$^{-6}$ mbar. ITO nanostructures of $\sim 200\ nm$ width and $\sim 500\ nm$ periodicity were patterned using electron beam lithography (Raith Pioneer 2) using $10\ wt\%$ PMMA resist. Lithographed samples were developed in a solution of methyl isobutyl ketone (MIBK) and isopropyl alcohol (IPA) in 1:3 volumetric ratio for 10 s followed by IPA wash for 10 s. Finally, $\sim 150\ nm$ ITO was sputter coated onto the patterned substrates with lift-off using warm acetone to yield the grating structure. Finally, the nanostructures samples were annealed using rapid thermal annealer (Mini Lamp Annealer-MILA 5000) for 15 minutes at 450°$C$ in $O_2$ lean atmosphere (~10$^{-6}$ mbar pressure), which yielded ITO with $\lambda_{ENZ} \sim 1220\ nm$.

**Material and optical Characterization**

Morphological characterization was conducted using Nova Nano SEM 450 field emission scanning electron microscope (SEM) coupled with an Apollo X energy dispersive x-ray analysis (EDS) system. X-ray diffraction (XRD) studies were performed with Cu $K_\alpha$ = 1.540 Å at 45 kV (Empyrean, PANanalytical). Raman spectra was obtained using a HORIBA Xplora Plus Raman setup with 532 nm laser excitation. Atomic force microscopy images were taken using a Bruker Multimode 8 AFM. Reflectance spectra were obtained in the range of $400 - 2500\ nm$ (Perkin Elmer Lamb da 950) using the Universal Reflectance Accessory module as well as the 60 mm Integrating Sphere module.

5. **Author's Contribution**

SD performed experiments and simulations. KPS and SD performed sputtering experiments. JM supervised the project. SD and JM analyzed data and wrote the manuscript with inputs from all the authors.

## 6. Acknowledgement

The authors thank Dr. S Kumaragurubaran, IISER TVM for the use of experimental facilities. SD acknowledges useful discussions with Dr. Ben Johns, IISER Mohali. The authors acknowledge ISTEM, Government of India, for access to the COMSOL Multiphysics software and financial support from SERB, Government of India (No. CRG/2023/006878). SD acknowledges a PhD fellowship from DST INSPIRE.

**Notes**

There are no conflicts to declare.

# Supporting information

**S1. Theoretical Modelling**

**S1.1 Model Details**

The finite element method is Maxwell's solver and a widely used numerical simulation approach for determining electromagnetic properties. A unit cell of the investigated system was designed in COMSOL 5.3a as shown in Fig.S1 and was simulated using periodic boundary conditions.

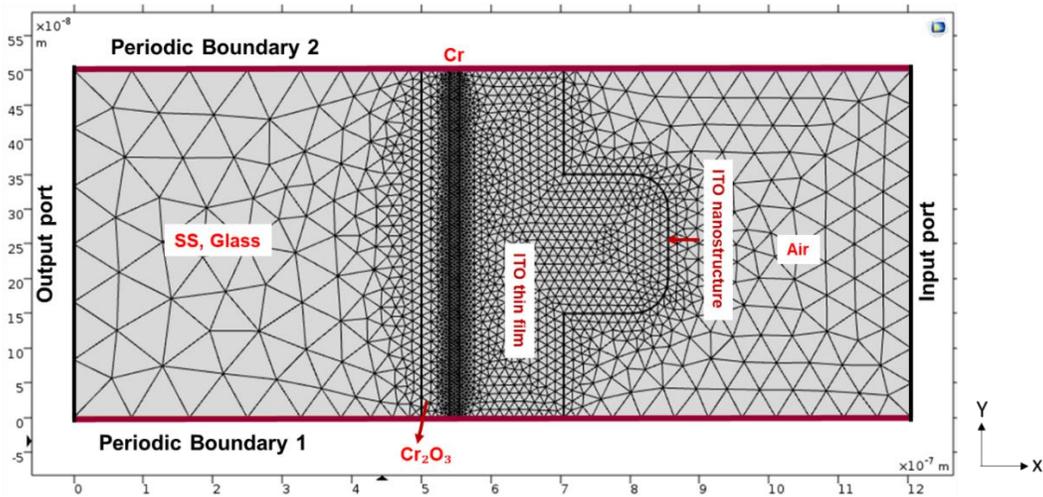

Figure S1: Cross sectional view of a unit cell designed in Wave optics module of COMSOL Multiphysics 5.3a

Two periodic ports were used where one was used to launch the incident wave and the other to collect the transmitted wave. Simulations were conducted using p-polarized light at selected angle of incidence and the reflectivity (R) was calculated. SS being a completely opaque substrate, the transmission through the system was zero.

**S1.2 Optimization details of the nanostructure**

The geometric dimensions of the nanostructures (NS) (width, height), as well as the periodicity of the array, were optimized systematically in COMSOL while developing the spectrally selective system. The width was varied keeping the height fixed, the height was varied keeping the width fixed and finally, the periodicity (centre to centre distance) was varied keeping the width and height fixed. Figure S2. shows that the nanostructure with ~200 nm width and ~150 nm height with ~500 nm periodicity helps to achieve the typical "step function" like reflectivity.

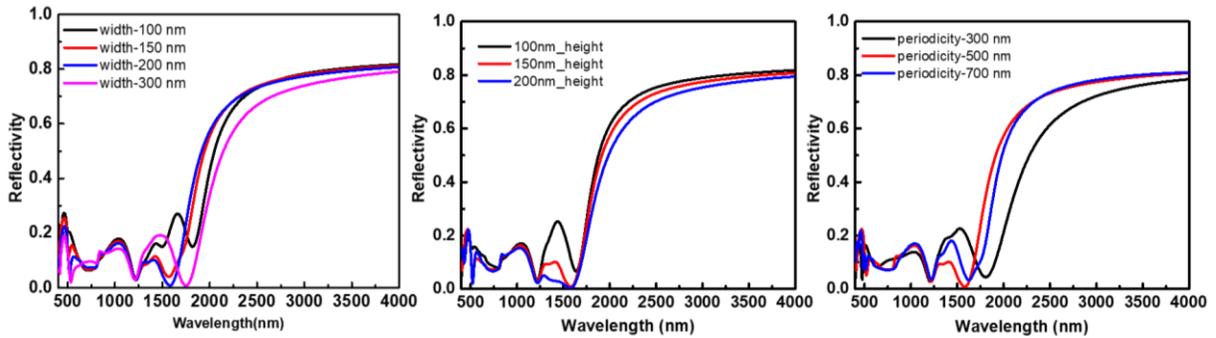

Figure S2(a) Simulated reflectivity plots for different width (100 nm, 150 nm, 200 nm, 300nm) of the NS, (b) different height (100 nm, 150 nm, 200 nm) of the NS, (c) different periodicity (300 nm, 500 nm, 700 nm) of the NS array.

## S2. Material and morphological characterization

### S2.1 Scanning electron microscopy (SEM)

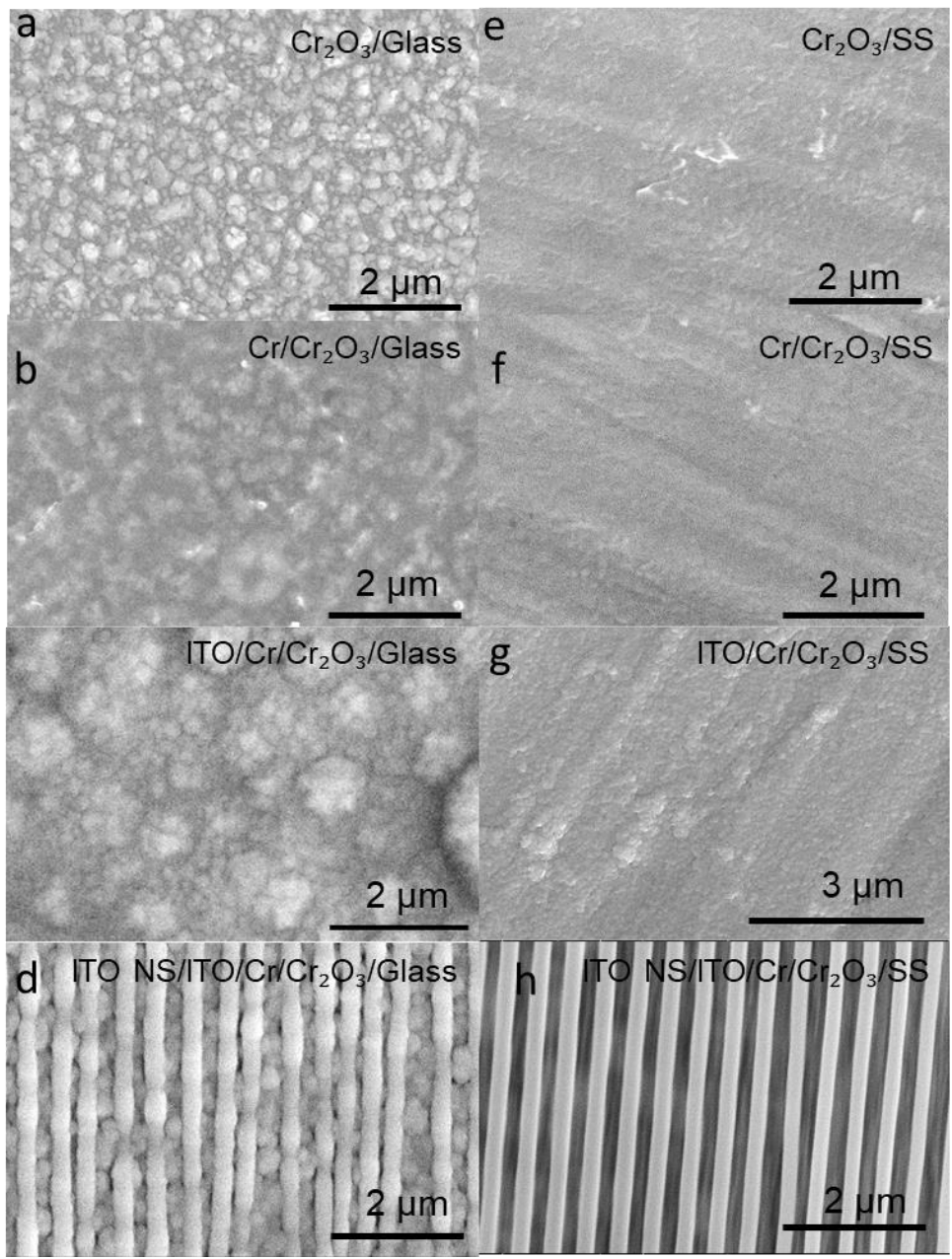

Figure S3. SEM images of the substrates after deposition of each layer, (a) $Cr_2O_3$, (b) $Cr/Cr_2O_3$ (c) $ITO/Cr/Cr_2O_3$ (d) ITO NS/ $ITO/Cr/Cr_2O_3$ on glass, (e) $Cr_2O_3$, f. $Cr/Cr_2O_3$, (g) $ITO/Cr/Cr_2O_3$, (h) ITO NS/ $ITO/Cr/Cr_2O_3$ on SS.

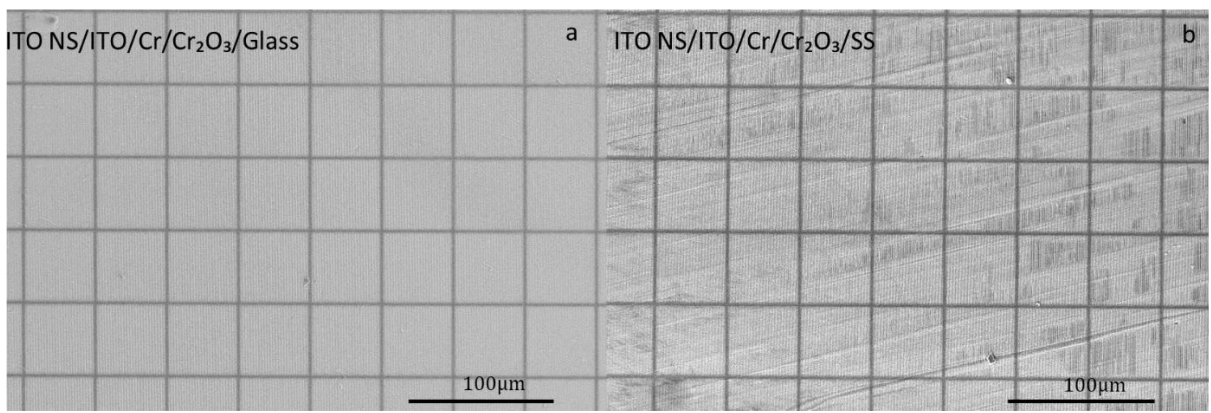

Figure S4. Large area SEM Images of the final multi-layered sample on (a) Glass substrate, (b) SS substrate

**S2.1 Energy dispersive spectroscopy (EDS), X-ray Diffraction (XRD) and Raman spectroscopy.**

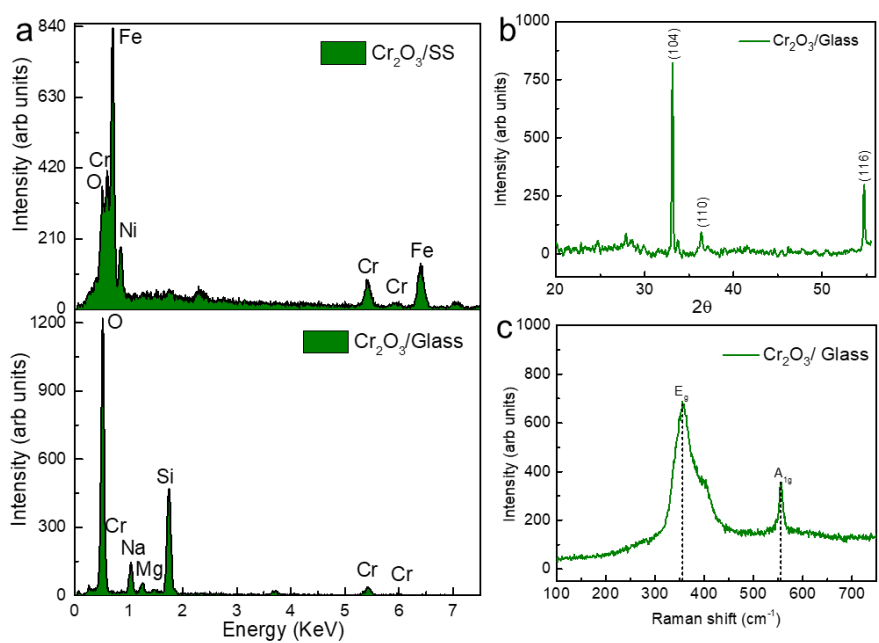

Figure S5 (a) EDS spectra of $Cr_2O_3$ on SS and glass, (b) XRD of $Cr_2O_3$ on glass, (c) Raman spectra of $Cr_2O_3$ on glass

## S2.3 Surface characterization using atomic force microscopy (AFM)

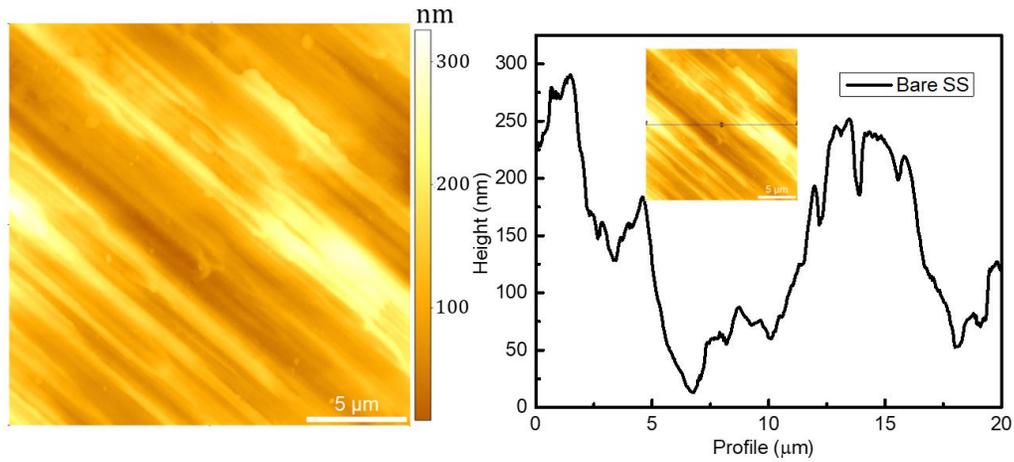

Figure S6. AFM topography of bare SS with the line-scan showing the height variation

The polished SS substrate shows significantly higher rms roughness (~ 120 nm) than the BK7 glass (roughness < 3 nm), estimated over $20 \times 20 \ \mu m^2$ areas. The average roughness of bare SS was calculated to be 118.44 nm

## S3. Transmission for the multilayer coating on glass substrate

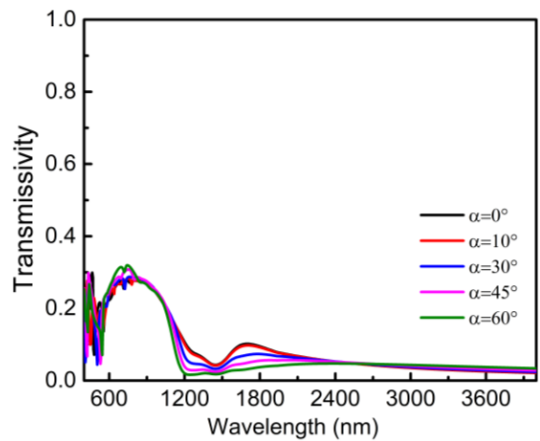

Figure S7. Transmissivity of the multilayer coating on glass substrate

## S4. Optical images of the final samples showing structural colours.

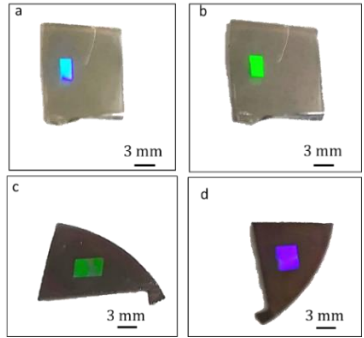

Figure S8. Optical images of the spectrally selective reflector on (a,b) glass substrate captured from different angles; (c,d) SS substrate captured from different angles. Different colours on the samples demarcate the nanostructured portion.

## S5. Nanostructure resonance

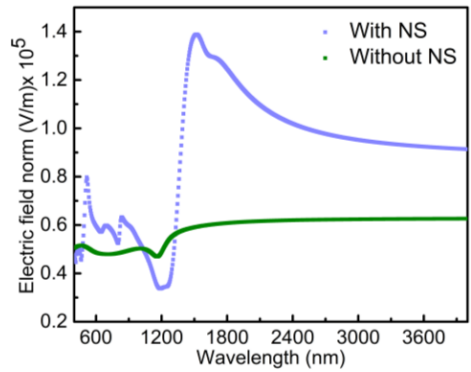

Figure S9. Simulated electric field magnitude (V/m) in the vicinity of NS and flat ITO without NS.

## S6. Tuning of $\lambda_{ENZ}$

| Annealing time (minutes) | 0 | 10 | 20 | 24 | 30 |
|---|---|---|---|---|---|
| $\lambda_{ENZ}$(nm) obtained | 1220 | 1255 | 1350 | 1450 | 1580 |

Table S1: Annealing of multilayer on SS in ambient pressure conditions

| Annealing time (minutes) | 0 | 10 | 20 | 30 | 40 |
|---|---|---|---|---|---|
| $\lambda_{ENZ}$(nm) obtained | 1220 | 1245 | 1310 | 1350 | 1450 |

Table S2: Annealing of multilayer on glass in ambient pressure conditions

## S7. Calculation of absorptivity (A) and Emissivity(E)

The absorptivity(A) and emissivity (E) of the coating on SS substrate was calculated using the equations 1 and 2 respectively.

$$A = \frac{\int_{0.4\,\mu m}^{1.5\,\mu m}(1 - R(\theta,\lambda))I_s(\lambda,\,T)\,d\lambda}{\int_{0.4\,\mu m}^{1.5\,\mu m} I_s(\lambda,\,T)\,d\lambda} \qquad [1]$$

$$E = \frac{\int_{2\,\mu m}^{4\,\mu m}(1 - R(\theta,\lambda))I_b(\lambda,\,T)\,d\lambda}{\int_{2\,\mu m}^{4\,\mu m} I_b(\lambda,\,T)\,d\lambda} \qquad [2]$$

$I_s$ is the solar radiation spectrum, $I_b$ is the blackbody radiation spectrum at 300 $K$ and $R$ is the calculated reflectivity at a particular angle $\theta$ and wavelength $\lambda$.

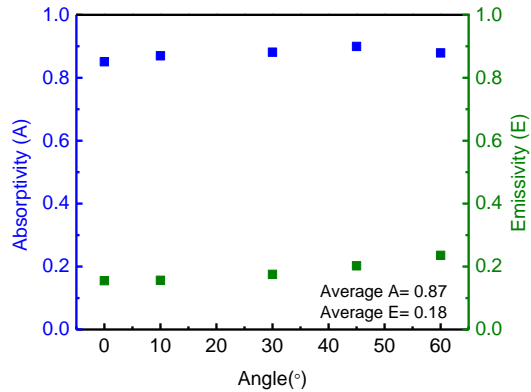

Fig S10. Variation of Calculated absorptivity and emissivity of multilayer on SS with angle

## S8. Role of the substrate

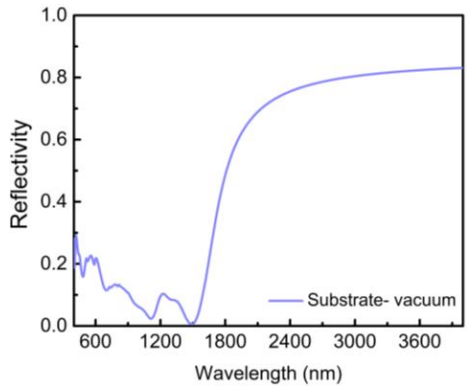

Fig S11. Simulated reflectivity of the spectrally selective coating at 45º with vacuum as the background substrate.